\documentclass[11pt,a4paper]{article}
\usepackage{jheppub_kim}

\usepackage{pdflscape}
\usepackage{amsmath}
\usepackage{amssymb}
\usepackage{dcolumn}
\usepackage{bm}
\usepackage{color}
\usepackage{epsfig}
\usepackage{amsfonts}
\usepackage{graphicx}
\usepackage{subfigure}
\usepackage{dcolumn}

\begin{document}

\title{Logarithmic corrected Van der Waals black holes in higher dimensional AdS space}

\author[a,b]{Sudhaker Upadhyay,}
\author[c]{and Behnam Pourhassan}

\affiliation[a]{Department of Physics, K.L.S. College,  Nawada-805110, Bihar, India}
\affiliation[b]{Visiting Associate, Inter-University Centre for Astronomy and Astrophysics (IUCAA) Pune, Maharashtra-411007}
 \affiliation[c]{School of Physics, Damghan University, Damghan, 3671641167, Iran}

\emailAdd{sudhakerupadhyay@gmail.com, sudhaker@associates.iucaa.in}
 \emailAdd{b.pourhassan@du.ac.ir}

\abstract{In this paper, we  consider AdS black hole in $d$-dimensional space-time and calculate the logarithmic corrected thermodynamics quantities. We study the effects of 
  thermal fluctuations on the thermodynamics of higher dimensional AdS black hole. We  exploit such logarithmic corrected quantities to construct Van der Walls black holes solution and study the effects of logarithmic correction on the Van der Waals black holes. We  also investigate the effect of thermal fluctuations on the equation of state. We find that the Van der Waals black hole is completely stable in the presence of the logarithmic correction.}


\maketitle

\section{Introduction}

Due to  AdS/CFT correspondence \cite{AdS-CFT}, it is possible to construct a holographic fluid dual to the black holes. It helps us to understand the strongly coupled thermal field theories living  on the boundary of AdS space-time. This idea has found several applications, for instance, the study  of the quark-gluon plasma \cite{Q1,Q2,Q3,Q4,Q5,Q6,Q7}. It has been found that the thermodynamics of an asymptotically AdS metric in four-dimensional space-time matches exactly with the thermodynamics of the van der Waals (VdW) fluid \cite{rajagopal}, which can be extended to the arbitrary number of dimensions and horizon topologies \cite{mann}. Therefore, the study of the black hole thermodynamics and their relation with the van der Waals fluid including charge or rotation is an interesting subject \cite{VdW1,VdW2,VdW3,VdW4,VdW5} of investigation in theoretical physics. One should note that such dual picture is   valid only if negative cosmological constant is considered as the thermodynamics pressure \cite{Ray,nc1,nc2,nc3,nc4,nc5,nc6,nc7,nc8}. 

The purpose of this paper is to study the effects of thermal fluctuations   on the above mentioned dual picture. The thermal fluctuations can be interpreted as quantum effect \cite{sudb1}. Since black hole is a gravitational system, hence the study of   quantum effect on such a system gives us information about quantum gravity which is  our primary motivation. The gravity may have correction near the Planck scale and such quantum gravity corrections  modify the topology of space-time at Planck scale \cite{6, 6a}.  The correction terms coming from quantum gravity can modify the entropy-area relation which is calculated by the holographic principle. The black hole entropy in absence of the correction terms is given by  $S_0= A/4$, where $A$ is the area of the black hole event horizon. The corrected relation of a black hole may be written as logarithmic term plus $S_0$. In that case the correction of the form $\ln A$  has  already been used to study the thermodynamics of the G\"{o}del black hole \cite{godel}. In fact the corrected thermodynamics of black holes can be studied by using the non-perturbative general relativity \cite{1z}. The usual way to study effects of quantum corrections to the black hole thermodynamics is the Cardy formula \cite{card}. We know that the matter fields around black holes may affect the thermodynamics \cite{other, other0, other1}. Also from the string theory poin of view the correction terms has been found in agreement with the other quantum gravity approaches \cite{solo1, solo2, solo4, solo5}.\\
The corrections to the thermodynamics of a dilatonic black hole have also been studied in the Ref. \cite{jy}. The black hole partition function is also useful to investigate the corrected thermodynamics of a given black hole \cite{bss}. As discussed by the Refs. \cite{mi, r1}, the generalized uncertainty principle can used to obtain the logarithmic correction. In that case the Einstein equations in the Jacobson formalism are realized as the thermodynamics identities \cite{z12j, jz12}. Therefore, the effect of a quantum correction to the space-time topology is a thermal fluctuations in the given black holes thermodynamics  \cite{l1, SPR, more}. Recently, the quantum corrections to thermodynamics of quasitopological black holes
have been studied \cite{sud17}. 

The mentioned corrections of the entropy already considered to study thermodynamics of several black objects. For example, a charged rotating AdS black holes \cite{sud18}, an AdS charged black hole \cite{1503.07418}, a Schwarzschild-Beltrami-de Sitter black hole \cite{sud}, a black saturn \cite{1505.02373}, a modified Hayward black hole \cite{1603.01457}, or a charged dilatonic black saturn \cite{1605.00924}. Also, the corrected Thermodynamics of a small singly spinning Kerr-AdS black hole has been studied by the Ref. \cite{NPB}. In that case, one may use results to study quantum gravity effects for example by using dumb holes \cite{Annals}. Similar analysis in case of the extremal black holes may be an important subject of theoretical physics \cite{ex1,ex2}. The first and second order corrections may affect the critical thermodynamics behaviors of some black object like a dyonic charged AdS black hole \cite{PRD}  and AdS black holes in massive gravity \cite{sudb}, which are considered as the holographic dual of a Van der Waals fluid to show that the holographic picture is still valid even for the logarithm-corrected thermodynamics. One of the important application of the logarithmic correction is to study the properties of the quark-gluon plasma  \cite{JHEP,CAN,EPJC2,G} like shear viscosity to entropy ratio \cite{EPJC}.\\

In this paper, we   consider AdS black hole in $d$ dimensional space-time and obtain corrected thermodynamics under effect of thermal fluctuations. Then, we use our results to obtain Van der Waals black hole and study the quantum effects on a holographic fluid.
In particular, we consider a higher dimensional AdS black hole with a metric
 given in terms of an unknown function $h(r, P_\Lambda)$. This
 function is determined by  the thermodynamics relations. We
derive the various corrected thermodynamics quantities
due to the first-order (logarithmic) correction in entropy. In particular,
we compute the first-order corrected  enthalpy, volume, Gibbs free
energy and, Helmholtz free energy and specific heat.
In order to derive the specific form of the unknown function $h(r, P_\Lambda)$,
we consider the logarithmic corrected  black hole equation of state (EoS)
and compare  with the corresponding
fluid equation of state by assuming equivalence between the black hole
and fluid temperatures, the black hole and fluid
volumes, and the cosmological and fluid pressures. The
  behavior of the function $h$ in terms of the model parameters is discussed with the help of
   plots.  We discuss the effects of thermal fluctuations on various
equation of states. For instance, we discuss the correction due to thermal fluctuations on
temperature, entropy,   mass,  volume and Gibbs free energy.  Finally, we  study the
stability of the model for  higher space-time dimensions, namely, $d=4,5$ and $d=10$.

The paper is organized as following. In the next section, we provide a metric ansatz
for the higher dimensional AdS black hole. In section 3, we discuss the thermodynamics
of the system having thermal fluctuations with logarithmic  term in the leading order.
The Van der Walls black hole is discussed in section 4. The effects due to the thermal
fluctuations are reported in section 5. In section 6, we study the stability of the model by observing sign of specific heat. We draw final remarks in the last section.
\section{Higher dimensional AdS black hole}
The line element for AdS Black Hole in $d$ dimensions is given by \cite{T},
\begin{eqnarray}\label{metric}
ds^2= -f(r)dt^2 +\frac{dr^2}{f(r)}+r^2 d\Omega^2_{d-2},
\end{eqnarray}
where $d\Omega^2_{d-2}$ is metric of the unit $d-2$ dimensional surface, and the function $f(r)$ is defined as,
\begin{equation}
f(r)=\frac{r^2}{l^2}-\frac{\mu}{r^{d-3}} -h(r, P_\Lambda),
\end{equation}
where $l$ is the cosmological radius related to the cosmological constant,
\begin{equation}
\Lambda=-\frac{(d-1)(d-2)}{2l^{2}}.
\end{equation}
Here $\mu$ is related to the AdM mass of the black hole which is given by,
\begin{equation}
M=\frac{(d-2)\Omega_{d-2}}{16\pi}\mu.
\end{equation}
The Lagrangian of the model considered such that the Riemann curvature not included
in matter Lagrangian.
We clarify that such a metric is a solution of the
Einstein field equations with a given energy-momentum source.
The energy-momentum source
would be physically  feasible if  it  satisfies certain conditions
such as positivity of energy density and dominance of the energy
density over pressure, known as energy conditions  \cite{po}.
Moreover, the unknown function $h(r, P_\Lambda)$ will be determined by using the thermodynamics relations. In order to have consistent solutions  with the asymptotic AdS structure, one can assume the following ansatz \cite{mann}:
\begin{equation}\label{ansatz}
h(r, P_\Lambda)=A(r)-P_{\Lambda}B(r).
\end{equation}
In the planar AdS black hole, $h=0$ and we have $r_{h}=(\mu l^{2})^{\frac{1}{d-1}}$. Generally, we have,
\begin{equation}
h(r_{h}, P_\Lambda)=\frac{r_{h}^{d-1}-\mu l^{2}}{l^{2}r_{h}^{d-3}}.
\end{equation}
Also, the momentum $V_{\Lambda}$ conjugated to $P_\Lambda$, which is interpreted as a volume, is given by
\begin{equation}\label{vol}
V_{\Lambda}=\frac{\partial M}{\partial P_{\Lambda}}.
\end{equation}
We will use the above relation to study the corrected thermodynamics of system in the next section.

\section{Logarithmic corrected thermodynamics}
It has been argued that the thermal fluctuations correct the thermodynamics quantities. To the leading order, these corrections are logarithmic   in the entropy expression \cite{l1, SPR}. The corrected entropy is given by,
\begin{eqnarray}\label{entropy}
S=S_0-\frac{\alpha}{2}\log(S_0T^2)+...,
\end{eqnarray}
where dots denote higher order corrections which are very small and can be neglected. The uncorrected entropy $S_0$ is given by
\begin{eqnarray}
S_0=\frac{\Omega_{d-2}}{4}r_h^{d-2}.
\end{eqnarray}
Also, the temperature is given by \cite{mann},
\begin{equation}\label{T}
T=\frac{f^{\prime}(r_h)}{4\pi} =\frac{1}{4\pi}\left[(d-1)\frac{r_h}{l^2}-
(d-3)\frac{h(r_h, P_\Lambda)}{r_h} -
 \frac{\partial h(r_h, P_\Lambda)}{\partial r_h} \right].
\end{equation}
By plugging the values of $S_0$ and $T$ into the equation (\ref{entropy}), we get,
 \begin{eqnarray}
S&=&\frac{\Omega_{d-2}}{4}r_h^{d-2}-\frac{\alpha}{2}\log \frac{\Omega_{d-2}}{64\pi^2}
-\frac{\alpha}{2}(d-2)\log r_h  \nonumber\\
&-&   {\alpha} \log
\left[(d-1)\frac{r_h}{l^2}-
(d-3)\frac{h(r_h, P_\Lambda)}{r_h} -
 \frac{\partial h(r_h, P_\Lambda)}{\partial r_h} \right],
\end{eqnarray}
The first law of thermodynamics is given by,
\begin{equation}
dM=TdS+V_\Lambda dP_\Lambda,
\end{equation}
where $M$ denotes to the enthalpy and has the following form \cite{jafarzade}:
\begin{eqnarray}
M&=&\frac{(d-2)}{16\pi}\Omega_{d-2}r_h^{d-3}\left(\frac{16\pi P_\Lambda r_h^2}{(d-1)(d-2)} -h(r_h, P_\Lambda)\right)\nonumber\\
&-&\frac{\alpha}{4\pi} \left[\frac{(d-1)r_h}{l^2}- (d-3)\frac{h(r_h, P_\Lambda)}{r_h}\right]\nonumber\\
&-&\frac{\alpha}{8\pi} (d-2)\left[\frac{ r_h}{l^2}- \frac{\mu}{r_h^{d-2}}-\frac{  h(r_h, P_\Lambda)}{  r_h}   \right].
\end{eqnarray}
The black hole volume is obtained by using the equation (\ref{vol}) as follow,
\begin{eqnarray}\label{V}
V_\Lambda =\frac{\Omega_{d-2} r_h^{d-1}}{(d-1)}-\frac{ (d-2)}{16\pi}\Omega_{d-2} r^{d-3}
\frac{\partial h (r, P_\Lambda)}{\partial P_\Lambda}+
\alpha\frac{(3d-8)}{8\pi r_h}\frac{\partial h (r, P_\Lambda)}{\partial P_\Lambda}.
\end{eqnarray}
In the case of positive $\frac{\partial h (r, P_\Lambda)}{\partial P_\Lambda}$ and positive $\alpha$ (and also $d\geq3$), we can see that  the logarithmic correction increases the black hole volume. Positive $\alpha$ is reasonable because its value is unit in many different models.
The logarithmic corrected Gibbs free energy, $G= M-TS$,  is calculated as,
\begin{eqnarray}
G&=&\frac{(d-2)}{16\pi}\Omega_{d-2}r_h^{d-3}\left(\frac{16\pi P_\Lambda r_h^2}{(d-1)(d-2)} -h(r_h, P_\Lambda)\right)\nonumber\\
&-&\frac{\Omega_{d-2}}{16\pi} \left[(d-1)\frac{r_h^{d-1}}{l^2}-(d-3) {h(r_h, P_\Lambda)}{r_h^{d-3}} -
r^{d-2} \frac{\partial h(r_h, P_\Lambda)}{\partial r_h} \right]\nonumber\\
&-&\frac{\alpha}{4\pi} \left[\frac{(d-1)r_h}{l^2}-(d-3)\frac{h(r_h, P_\Lambda)}{r_h}-\frac{\partial h(r_h, P_\Lambda)}{\partial r_h} \right]\nonumber\\
&-&\frac{\alpha}{8\pi} (d-2)\left[\frac{ r_h}{l^2}- \frac{\mu}{r_h^{d-2}} -  \frac{  h(r_h, P_\Lambda)}{  r_h}   \right]\nonumber\\
&+& \frac{\alpha}{8\pi}\log \frac{\Omega_{d-2}}{64\pi^2}\left[(d-1)\frac{r_h}{l^2}-
(d-3)\frac{h(r_h, P_\Lambda)}{r_h} -
 \frac{\partial h(r_h, P_\Lambda)}{\partial r_h} \right]\nonumber\\
&+&\frac{\alpha}{8\pi}(d-2)\left[(d-1)\frac{r_h\log r_h }{l^2}-
(d-3)h(r_h, P_\Lambda)\frac{\log r_h }{r_h} -\log r_h
 \frac{\partial h(r_h, P_\Lambda)}{\partial r_h} \right]\nonumber\\
&+&  \frac{\alpha}{4\pi} \left[(d-1)\frac{r_h}{l^2}-
(d-3)\frac{h(r_h, P_\Lambda)}{r_h} -
 \frac{\partial h(r_h, P_\Lambda)}{\partial r_h} \right]\nonumber\\
&\times& \log\left[(d-1)\frac{r_h}{l^2} -(d-3)\frac{h(r_h, P_\Lambda)}{r_h} -
 \frac{\partial h(r_h, P_\Lambda)}{\partial r_h}  \right]
\end{eqnarray}
The pressure $P_\Lambda$ is given by \cite{mann},
\begin{equation}\label{P}
P_\Lambda =-\frac{\Lambda}{8\pi}=\frac{(d-1)(d-2)}{16\pi l^2},
\end{equation}
Hence, the Helmholtz free energy, $F= G-P_\Lambda V_\Lambda$,  is calculated by
\begin{eqnarray}
F& =&\frac{(d-2)}{16\pi}\Omega_{d-2}r_h^{d-3}\left(\frac{16\pi P_\Lambda r_h^2}{(d-1)(d-2)} -h(r_h, P_\Lambda)\right)\nonumber\\
&-&\frac{\Omega_{d-2}}{16\pi} \left[(d-1)\frac{r_h^{d-1}}{l^2}-(d-3) {h(r_h, P_\Lambda)}{r_h^{d-3}} - r^{d-2} \frac{\partial h(r_h, P_\Lambda)}{\partial r_h} \right]\nonumber\\
&-&\frac{(d-2)}{16\pi l^2}\Omega_{d-2} r_h^{d-1}+\frac{ (d-1)(d-2)^2}{(16\pi l)^2}\Omega_{d-2} r^{d-3}
\frac{\partial h(r, P_\Lambda) }{\partial P_\Lambda}\nonumber\\
&-&\frac{\alpha}{4\pi} \left[\frac{(d-1)r_h}{l^2}- (d-3)\frac{h(r_h, P_\Lambda)}{r_h}
-  \frac{\partial h(r_h, P_\Lambda)}{\partial r_h} \right]\nonumber\\
&-&\frac{\alpha}{8\pi} (d-2)\left[\frac{ r_h}{l^2}- \frac{\mu}{r_h^{d-2}} -  \frac{  h(r_h, P_\Lambda)}{  r_h}   \right]
\nonumber\\
&+& \frac{\alpha}{8\pi}\log \frac{\Omega_{d-2}}{64\pi^2}\left[(d-1)\frac{r_h}{l^2}-
(d-3)\frac{h(r_h, P_\Lambda)}{r_h} -
 \frac{\partial h(r_h, P_\Lambda)}{\partial r_h} \right]\nonumber\\
&+&\frac{\alpha}{8\pi}(d-2)\left[(d-1)\frac{r_h\log r_h }{l^2}-
(d-3)h(r_h, P_\Lambda)\frac{\log r_h }{r_h} -
 \frac{\partial h(r_h, P_\Lambda)}{\partial r_h} \log r_h  \right]\nonumber\\
 &-&\alpha\frac{(d-1)(d-2)(3d-8)}{128\pi^2 l^2 r_h}\frac{\partial h(r, P_\Lambda) }{\partial P_\Lambda}\nonumber\\
&+&  \frac{\alpha}{4\pi} \left[(d-1)\frac{r_h}{l^2}-
(d-3)\frac{h(r_h, P_\Lambda)}{r_h} -
 \frac{\partial h(r_h, P_\Lambda)}{\partial r_h} \right]\nonumber\\
&\times&\log\left[(d-1)\frac{r_h}{l^2}-(d-3)\frac{h(r_h, P_\Lambda)}{r_h} -
 \frac{\partial h(r_h, P_\Lambda)}{\partial r_h} \right].
\end{eqnarray}
The specific heat is defined by,
\begin{equation}
C=\left(\frac{\partial M}{\partial T}\right)_V.
\end{equation}
Using the above definition, we calculate the specific heat as,
\begin{eqnarray}\label{C}
C&=&\frac{\Omega_{d-2}\left[(d-1)\frac{r_h^{d-2}}{l^2} -(d-3) {h(r_h, P_\Lambda)}{r_h^{d-4}} -  r_h^{d-3}\frac{\partial h(r_h, P_\Lambda)}{\partial r_h} \right]}
{4\left[\frac{(d-1)}{l^2}+(d-3)\frac{h(r_h, P_\Lambda)}{r^2_h}-\frac{(d-3)}{r_h} \frac{\partial  h(r_h, P_\Lambda)}{\partial r_h}- \frac{\partial^2 h(r_h, P_\Lambda)}{\partial r_h^2} \right]}\nonumber\\
&-&\frac{\alpha\left[\frac{d(d-1)}{l^2} -(d-3)(d-4)\frac{ h(r_h, P_\Lambda)}{r_h^2} -\frac{(3d-5)}{r_h}\frac{\partial h(r_h, P_\Lambda)}{\partial r_h}- 2\frac{\partial^2 h(r_h, P_\Lambda)}{\partial r_h^2} \right]}
{2\left[\frac{(d-1)}{l^2}+(d-3)\frac{h(r_h, P_\Lambda)}{r^2_h}-\frac{(d-3)}{r_h} \frac{\partial  h(r_h, P_\Lambda)}{\partial r_h}- \frac{\partial^2 h(r_h, P_\Lambda)}{\partial r_h^2} \right]}.
 \end{eqnarray}
This expression can be used to investigate stability of the black hole with the condition $C\geq0$. In order to do that the form of function $h$ must be determined. We will do that in the next section, then try to study stability of the black hole.

\section{Van der Waals black holes}
\begin{figure}[h!]
 \begin{center}$
 \begin{array}{cccc}
\includegraphics[width=60 mm]{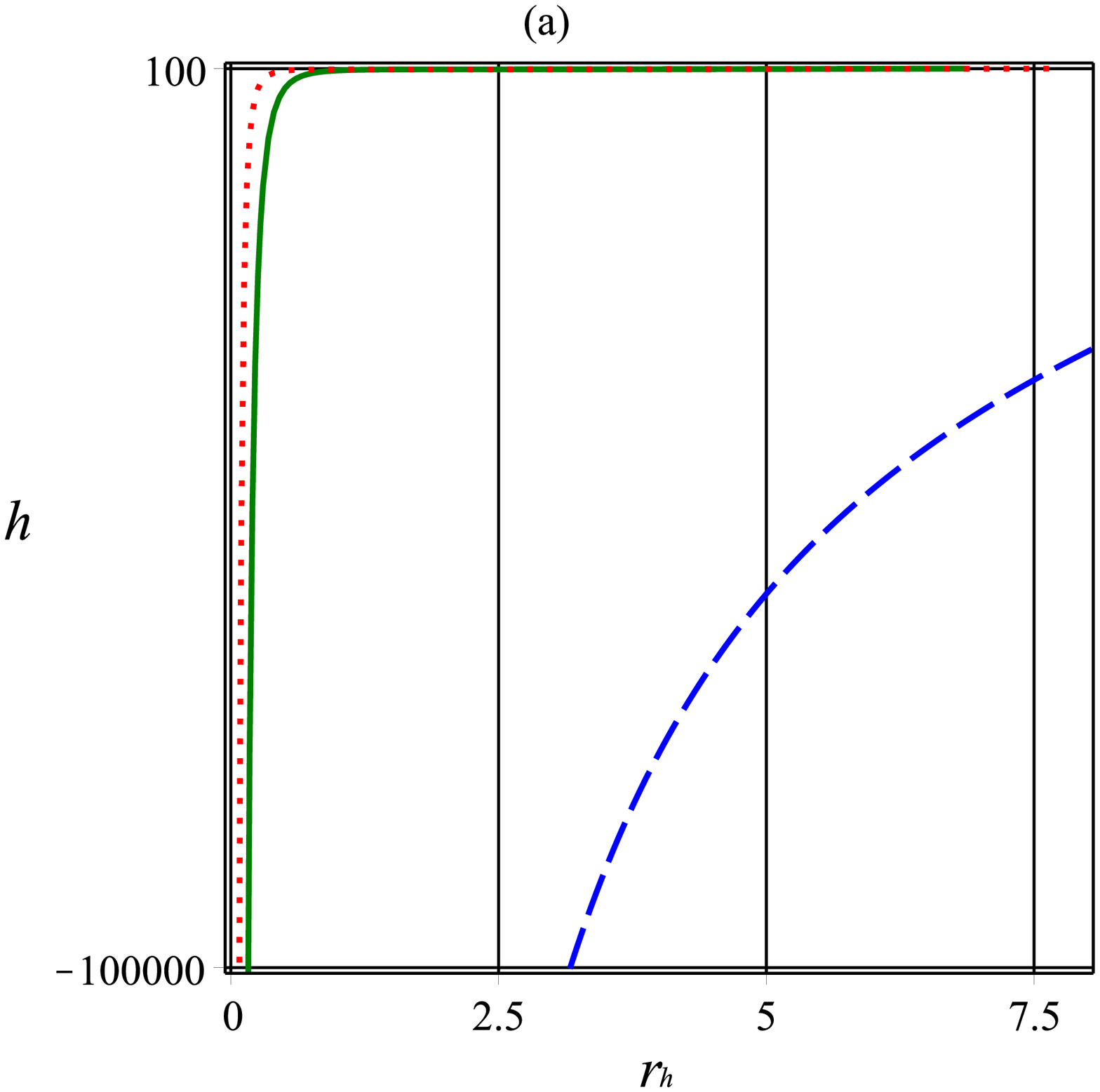}&\includegraphics[width=60 mm]{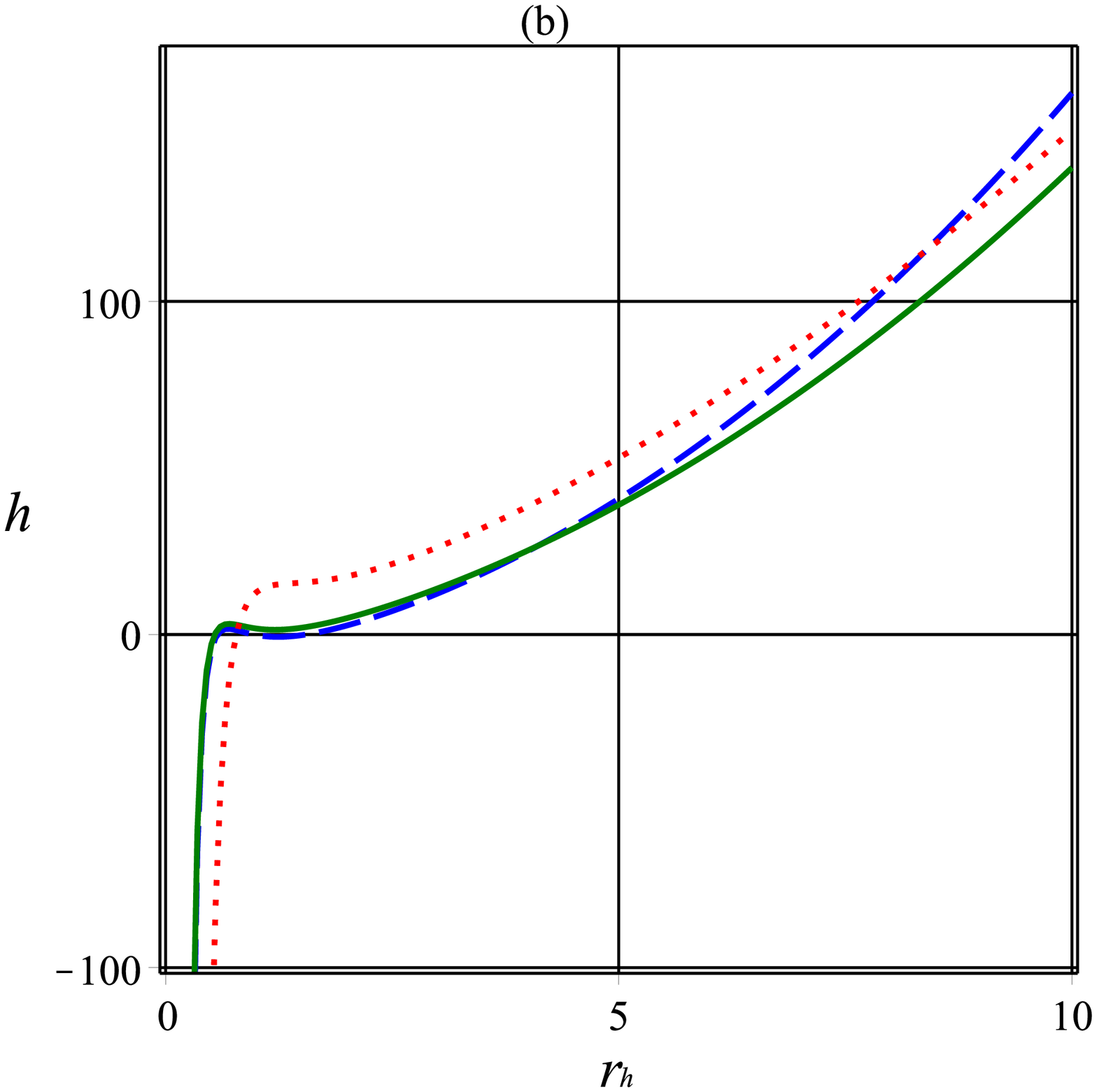}\\
\includegraphics[width=60 mm]{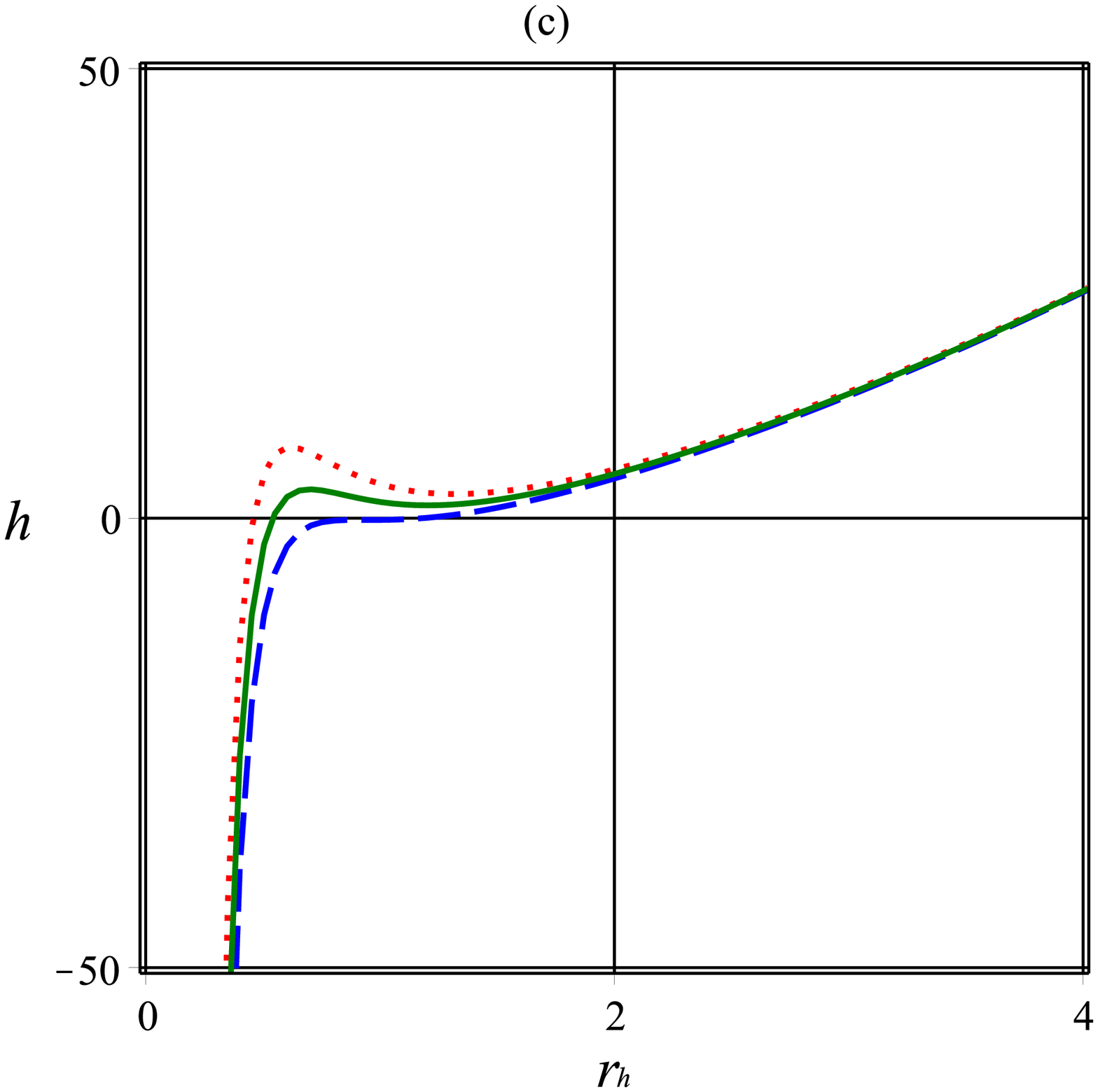}&\includegraphics[width=60 mm]{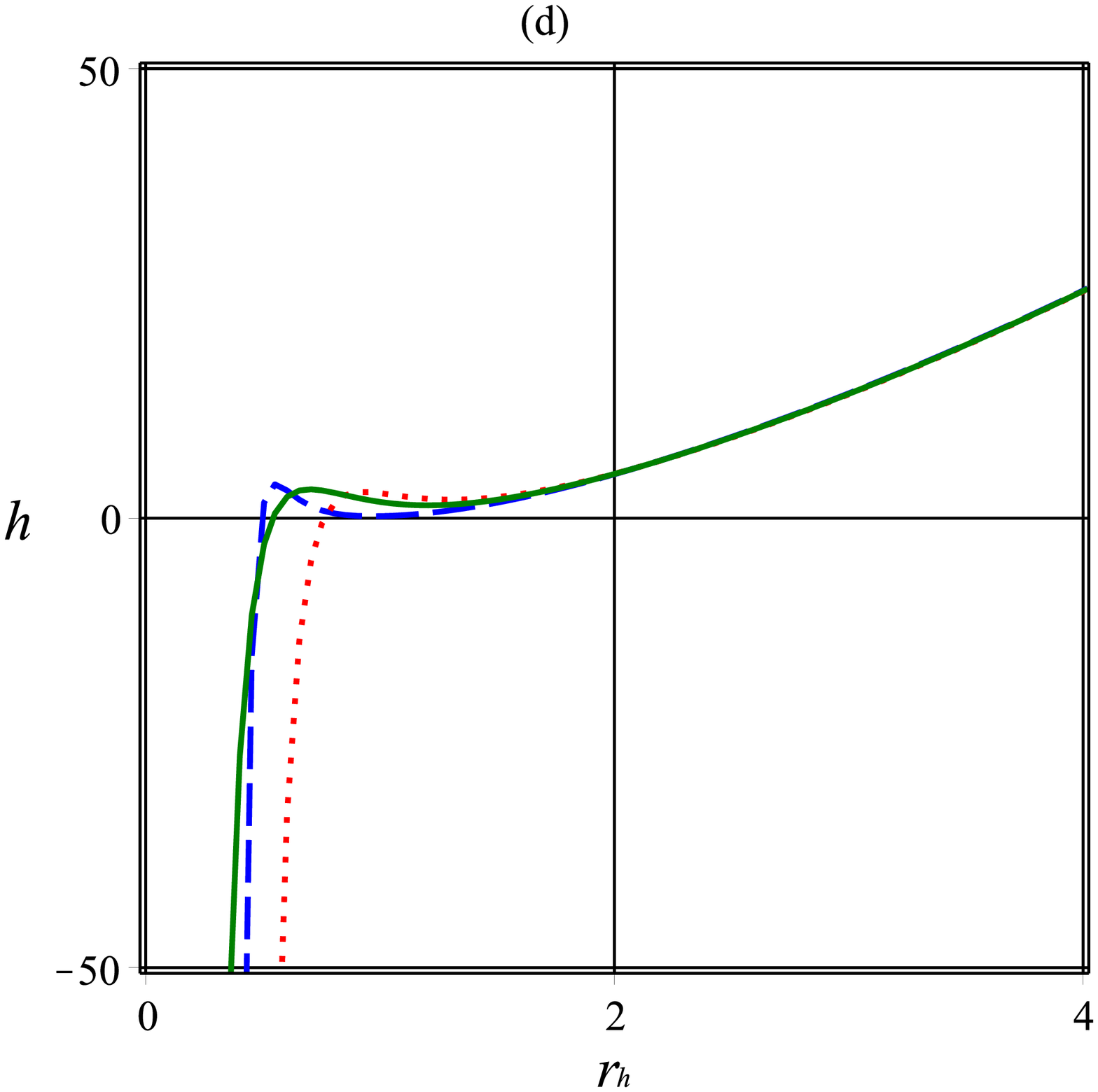}
 \end{array}$
 \end{center}
\caption{Typical behavior of $h$ in terms of $r_{h}$ by variation of the black hole parameters with $a=b=l=1$. (a) $d=4$, $C_l=C_M=0$; $\alpha=0.01$ (blue dash), $\alpha=0.5$ (green solid), $\alpha=1$ (red dot). (b) $\alpha=1$, $C_l=C_M=0$; $d=4$ (blue dash), $d=5$ (green solid), $d=10$ (red dot). (c) $d=5$, $\alpha=1$, $C_l=0$; $C_M=-2$ (blue dash), $C_M=0$ (green solid), $C_M=2$ (red dot). (d) $d=5$, $\alpha=1$, $C_M=0$; $C_l=-2$ (blue dash), $C_l=0$ (green solid), $C_l=2$ (red dot).}
 \label{fig1}
\end{figure}

In order to be consistent with the Van der Waals fluid,   the black holes must satisfy the
certain energy conditions  near the horizon, and mass should remain positive  (see details in \cite{rajagopal}).
A Van der Waals fluid is described by the following equation of state:
\begin{equation}\label{EoS}
T=\left(P+\frac{a}{V^{2}}\right)(V-b),
\end{equation}
where the constant $a$ parameterizes the value of the intermolecular interactions, and the constant $b$ consider for the volume of molecules. The ideal gas equation of state obtained by setting $a=b=0$. The equation of state (\ref{EoS}) is valid also in higher dimensions \cite{nc2}. In order to have a Van der Waals black hole we set $P=P_{\Lambda}$ and $V=v_{\Lambda}$ (here $v_{\Lambda}$ is specific volume given by $4\frac{(d-1)V_{\Lambda}}{(d-2)r^{d-2}}$) and used the equations (\ref{T}), (\ref{V}), and (\ref{P}) together with ansatz (\ref{ansatz}). This gives  the following differential equations for $A$ and $B$:
\begin{equation}\label{B}
r_{h}B^{\prime}+\frac{3}{2}\left(\alpha(d-\frac{8}{3})-\frac{4}{3}\right)B-4\pi b r_{h}+\frac{16\pi r_{h}^{2}}{d-2}=0,
\end{equation}
and
\begin{equation}\label{A}
r_{h}A^{\prime}+(d-3)A-\frac{d-1}{l^2}r_{h}^{2}+\frac{\frac{4r_{h}}{d-2}-\frac{(d-1)B}{4\pi r_{h}}+\frac{\alpha(3d-8)B}{8\pi r_{h}}-b}{\frac{4r_{h}}{d-2}-\frac{(d-1)B}{4\pi r_{h}}+\frac{\alpha(3d-8)B}{8\pi r_{h}}}4\pi a r_{h}=0,
\end{equation}
where prime denotes derivative with respect to $r_{h}$. It is easy to check the equation (\ref{B}) at $\alpha=0$ limit and $b\gg r_{h}$ coincides with the Ref. \cite{mann}. Now, general solution of the equation (\ref{B}) is,
\begin{equation}\label{B-1}
B=\frac{8\pi r_{h}\left[\alpha(d(3d-14)+16)b-4r_{h}(\alpha(3d-8)-2)\right]}{\alpha(3d-8)(d-2)\left(\alpha(3d-8)-2\right)}+C_{l}r_{h}^{-\frac{3}{2}\alpha(d-\frac{8}{3})+2},
\end{equation}
where $C_{l}$ is an integration constant.\\
Solution of the equation (\ref{A}), in the case of $\alpha=0$, is given in terms of the hypergeometric function \cite{mann}. Here, we obtain the following approximate solution for the equation (\ref{A}):
\begin{equation}\label{A-1}
A=A_{0}\left[\frac{C_{1}}{r_{h}}+\frac{C_{2}}{r_{h}^{2}}+\frac{C_{3}}{r_{h}^{3}}+\frac{C_{4}}{r_{h}^{\alpha(3d-8)+1}}+\frac{C_{5}}{r_{h}^{\frac{\alpha}{2}(3d-8)+1}}
+\frac{C_{6}}{r_{h}^{\frac{\alpha}{2}(3d-8)+2}}+\frac{C(r_{h})}{r_{h}^{d-3}}\right],
\end{equation}
where
\begin{equation}\label{A-C1}
C_{1}=-\frac{8}{9}\pi^{2}l^{2}a\left(\alpha(d-\frac{8}{3})-\frac{2}{3}\right)^{2}(d-2)(d-1)^{2},
\end{equation}
\begin{equation}\label{A-C2}
C_{2}=-\frac{1}{6}\alpha\pi^{2}l^{2}ab\left(\alpha(d-\frac{8}{3})-\frac{8d}{3}+4\right)(d-2)^{2}(d-1)(d-\frac{8}{3})\left(\alpha(d-\frac{8}{3})-\frac{2}{3}\right),
\end{equation}
\begin{equation}\label{A-C3}
C_{3}=\frac{1}{24}\alpha^{2}\pi^{2}l^{2}ab^{2}\left(\alpha(d-\frac{8}{3})-\frac{4d}{3}+\frac{4}{3}\right)(d-2)^{4}(d-\frac{8}{3})^{2},
\end{equation}
\begin{equation}\label{A-C4}
C_{4}=-\frac{9(d-2)^{3}}{1024}\alpha^{2}l^{2}aC_{l}^{2}\left(\alpha(d-\frac{8}{3})-\frac{4d}{3}+\frac{4}{3}\right)
\left(\alpha(d-\frac{8}{3})-\frac{2d}{3}+\frac{2}{3}\right)\left(\alpha(d-\frac{8}{3})-\frac{2}{3}\right)^{2},
\end{equation}
\begin{equation}\label{A-C5}
C_{5}=-\frac{3}{16}\pi\alpha l^{2}aC_{l}\left(\alpha(d-\frac{8}{3})-\frac{8d}{9}+\frac{8}{9}\right)(d-2)^{2}(d-1)
\left(\alpha(d-\frac{8}{3})-\frac{2}{3}\right)^{2},
\end{equation}
\begin{eqnarray}\label{A-C6}
C_{6}&=&-\frac{3(d-2)^{3}(d-\frac{8}{3})^{2}}{128}\pi\alpha^{2}l^{2}abC_{l}\nonumber\\
&\times&\left(\alpha(d-\frac{8}{3})-\frac{4d}{3}+2\right)
\left(\alpha(d-\frac{8}{3})-\frac{4d}{3}+\frac{4}{3}\right)\left(\alpha(d-\frac{8}{3})-\frac{2}{3}\right),
\end{eqnarray}
\begin{equation}\label{A-0}
A_{0}=-\frac{81}{\left(\pi\alpha^{2}l^{2}(3d-8)^{2}\left(\alpha(3d-8)-2\right)^{2}\right)^{2}},
\end{equation}
and
\begin{equation}\label{A-Cr}
C(r_{h})=\pi\alpha^{2}(\alpha(d-\frac{8}{3})-\frac{2}{3})^{2}(d-\frac{8}{3})^{2}\left((d-1)r_{h}^{d-2}+C_{M}l^{2}\right),
\end{equation}
where $C_{M}$ is another integration constant. In the case of $d=4$ and $\alpha=0$, it has been argued that $C_{l}=0$ together $C_{M}=3ab\pi$ recovers the results of the Ref. \cite{rajagopal}.
We can see that both $A$ and $B$ depend on space dimensions ($d$) as well as logarithmic correction parameter ($\alpha$), while in the case of $\alpha=0$ it has been found that $B$ is independent of $d$ \cite{mann}. Also, in the case of $C_{l}=0$, the cosmological pressure $P_{\Lambda}$,interpreted as the thermodynamics pressure, is demanded to be produced entirely by the fluid, which  increases the value of the effective AdS length.
Using above solutions, the equation of state (\ref{EoS}) get satisfied and the black hole described by the metric (\ref{metric}) becomes dual of a Van der Waals fluid.
In the Fig. \ref{fig1} we can see typical behavior of the function $h$ in terms of the model parameters.

From the Fig. \ref{fig1} (a), we see that the function $h$ increases sharply with respect to small $r_h$ and becomes
saturated  with $r_h$ for larger $\alpha$ (of the order of unit). However, as $\alpha$ tends to very small value, the function $h$ starts  increasing  rather slowly after a certain value of $r_h$.   The behavior of the function $h$ with respect to $r_h$ can be seen in Fig. \ref{fig1} (b) for
different space dimensions $(d)$. Here, we see that as dimension decreases
the  $h$ gets higher value as long as $r_h$ tends to higher value. The opposite behavior
occurs for small $r_h$ and as long as $r_h$ tends to zero the value of $h$ decreases
  sharply to negative values. The behavior of function $h$ with respect to $r_h$
  can be seen for different values of integration constants $C_M$ and $C_l$
  in Figs. \ref{fig1} (c) and \ref{fig1} (d) respectively. The peak value of $h$ increases
 with increasing  $C_M$ near small $r_h$ otherwise  the $C_M$ does not affect $h$
 with respect to $r_h$. Similar behavior is observed for $C_l$ also and $C_l$
 does not affect  $h$ for higher $r_h$.\\
Now, we can investigate the effect of logarithmic correction, coming from thermal fluctuations, on the thermodynamic quantities which obtained in the section 3.

\section{Effect of thermal fluctuations}
\begin{figure}[h!]
 \begin{center}$
 \begin{array}{cccc}
\includegraphics[width=60 mm]{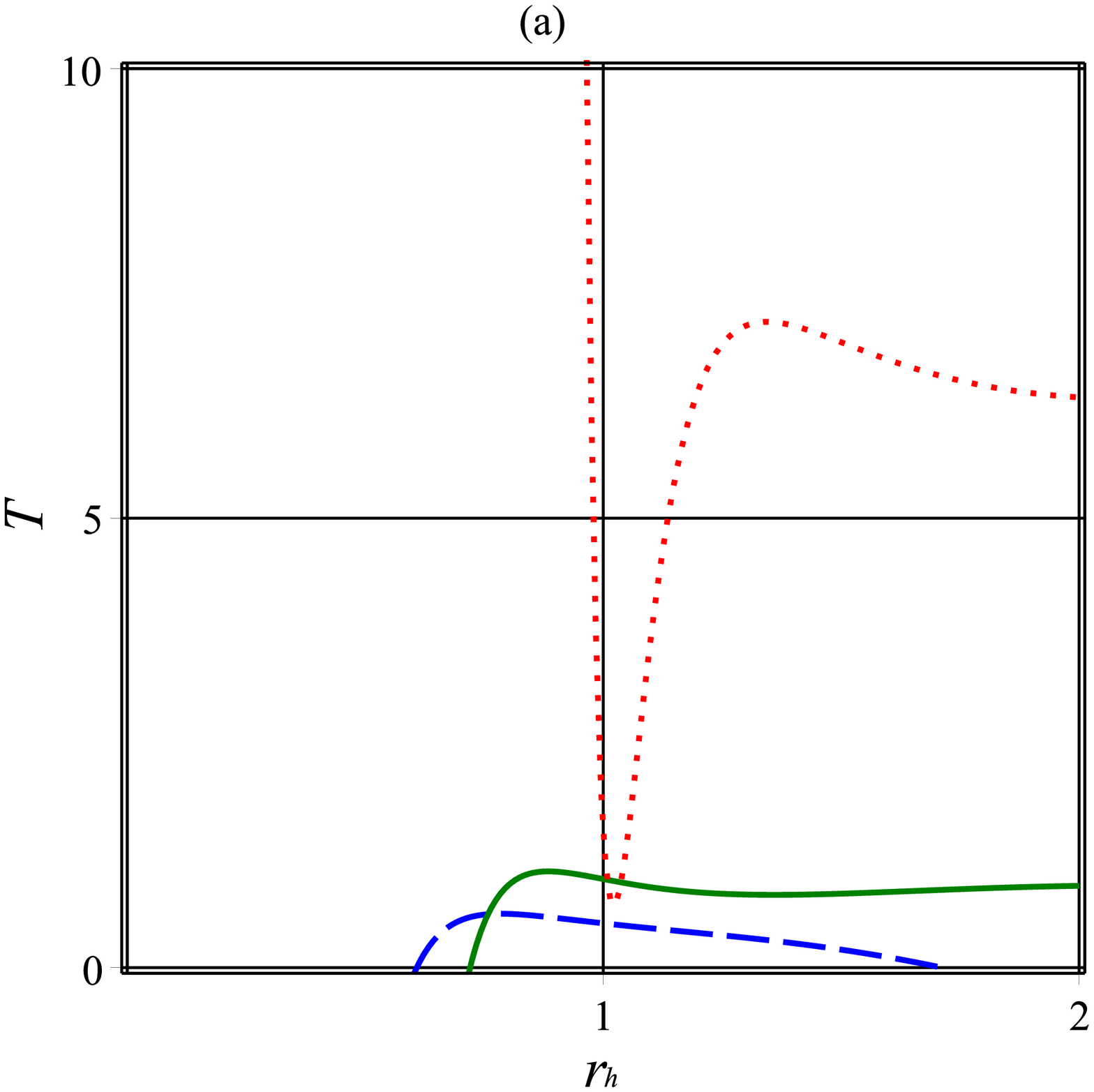}&\includegraphics[width=60 mm]{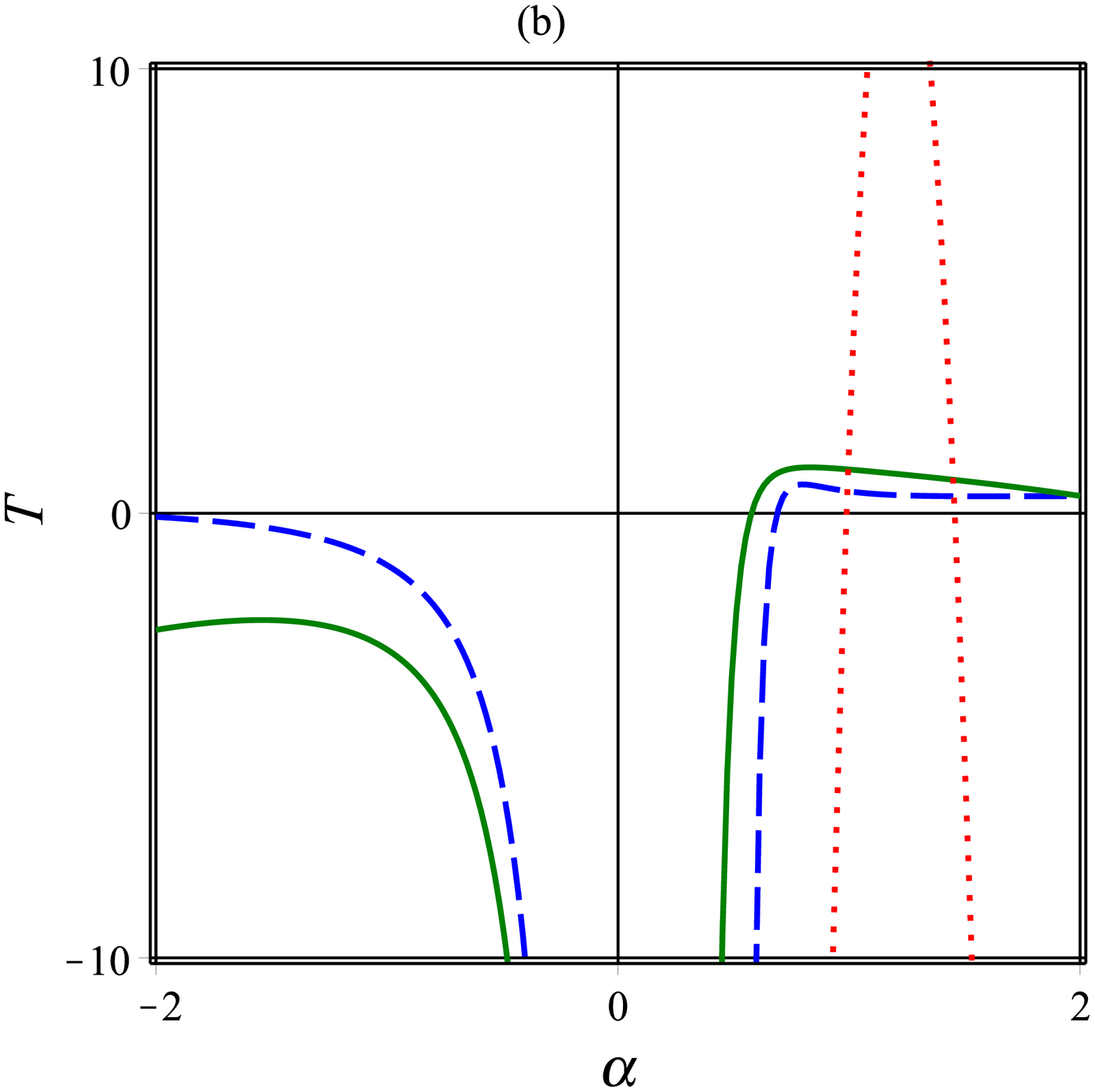}
 \end{array}$
 \end{center}
\caption{Typical behavior of temperature in terms of $r_{h}$ (a) and $\alpha$ (b) by variation of the space-time dimensions. We set the black hole parameters as $a=b=l=r_{h}=1$, $C_l=C_M=2$. $d=4$ (blue dash), $d=5$ (green solid), $d=10$ (red dot).}
 \label{fig2}
\end{figure}

\begin{figure}[h!]
 \begin{center}$
 \begin{array}{cccc}
\includegraphics[width=60 mm]{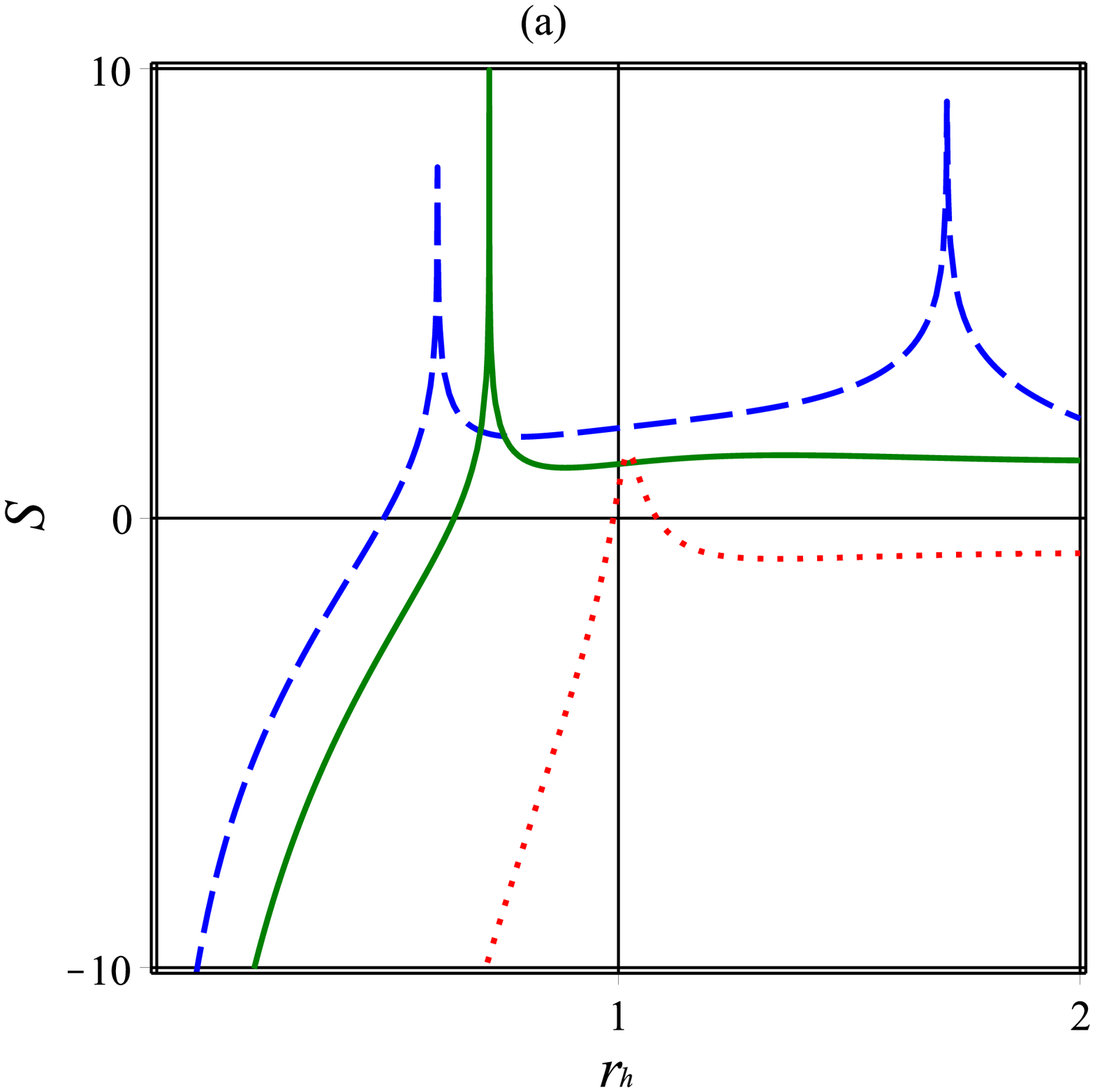}&\includegraphics[width=60 mm]{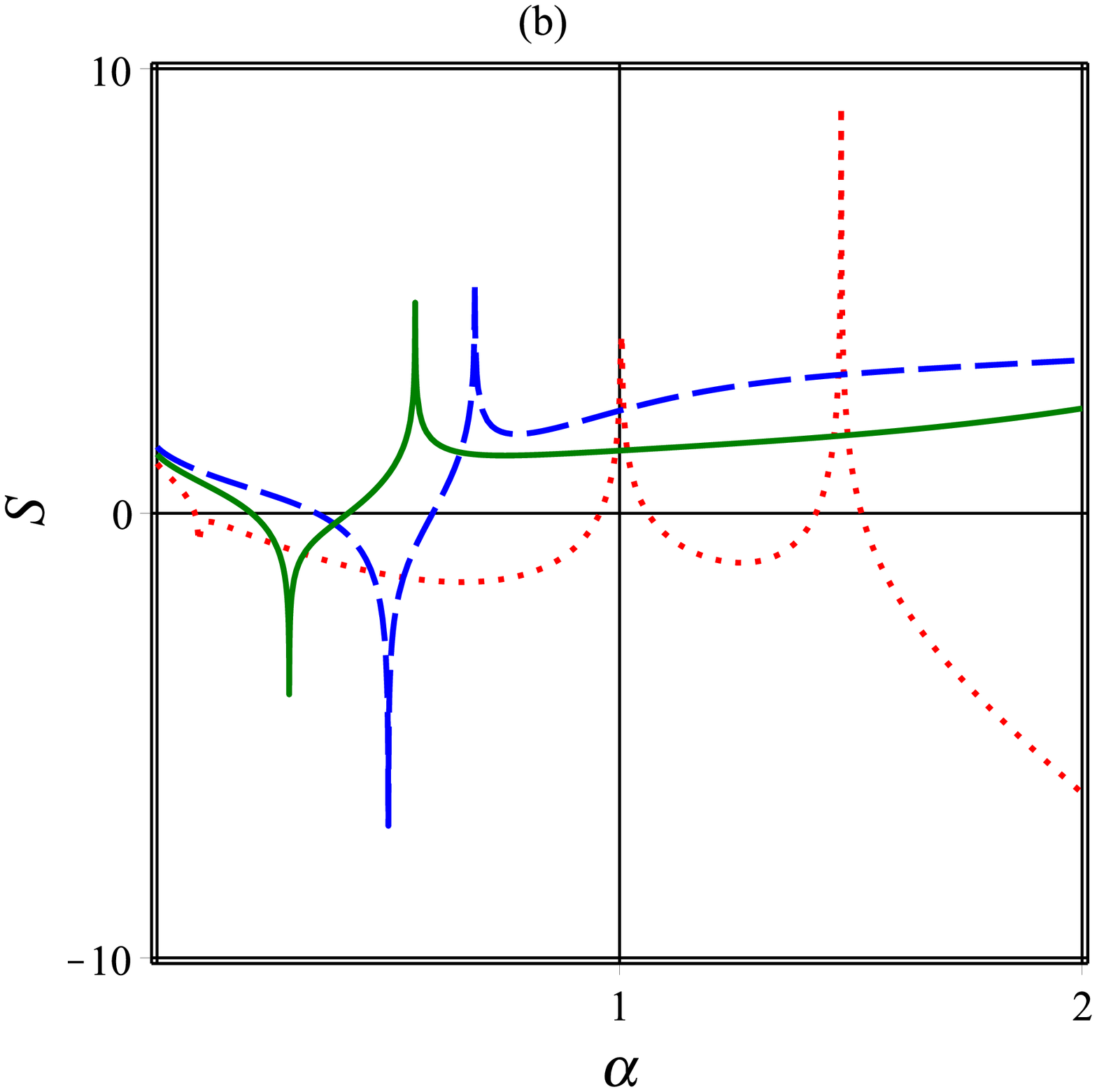}
 \end{array}$
 \end{center}
\caption{Typical behavior of the entropy in terms of (a) $r_{h}$ with $\alpha=1$, and (b) $\alpha$ with $r_{h}=1$, by variation of the space-time dimensions. We set the black hole parameters as $a=b=l=1$, $C_l=C_M=2$. $d=4$ (blue dash), $d=5$ (green solid), $d=10$ (red dot).}
 \label{fig3}
\end{figure}

\begin{figure}[h!]
 \begin{center}$
 \begin{array}{cccc}
\includegraphics[width=60 mm]{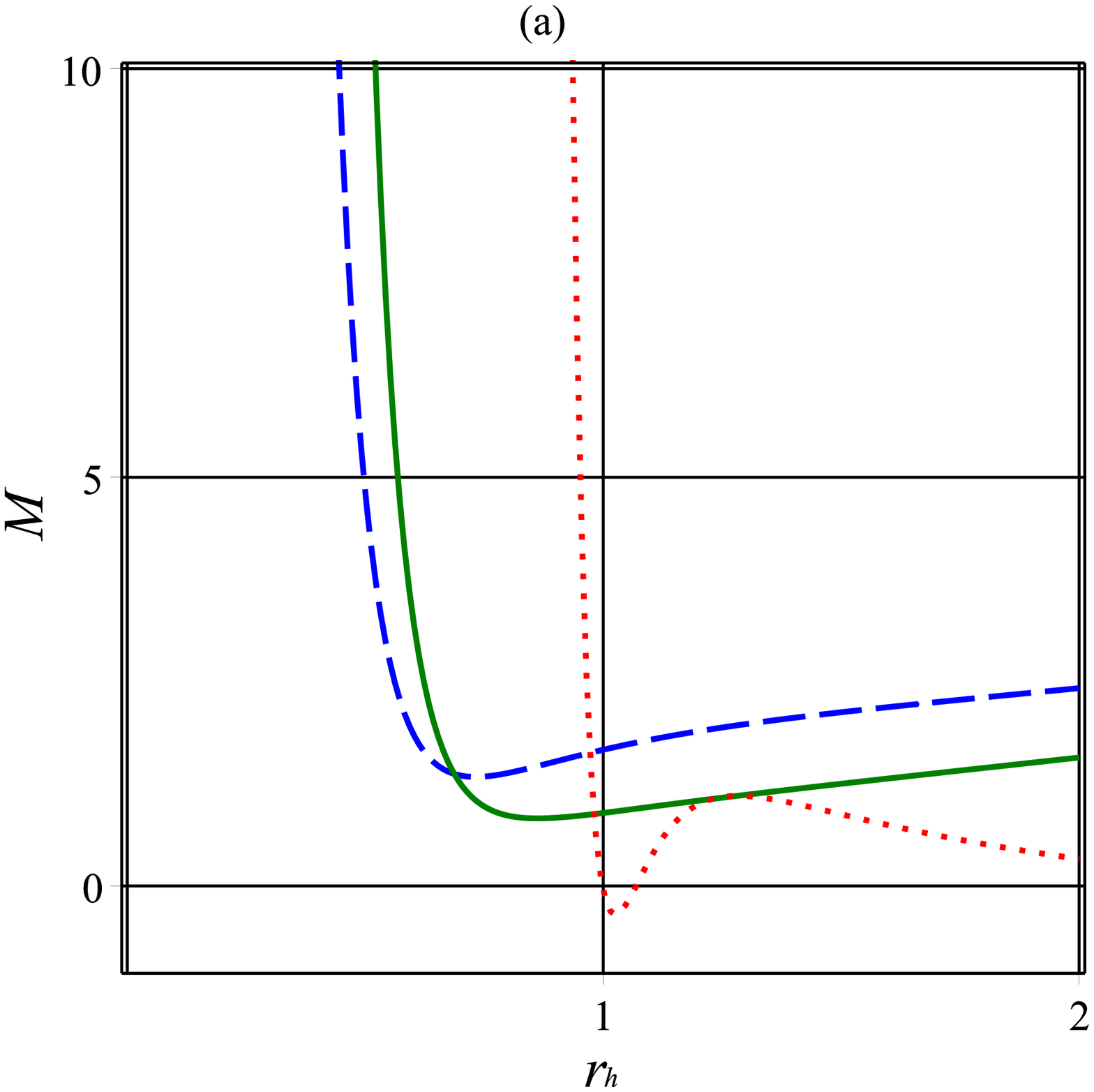}&\includegraphics[width=60 mm]{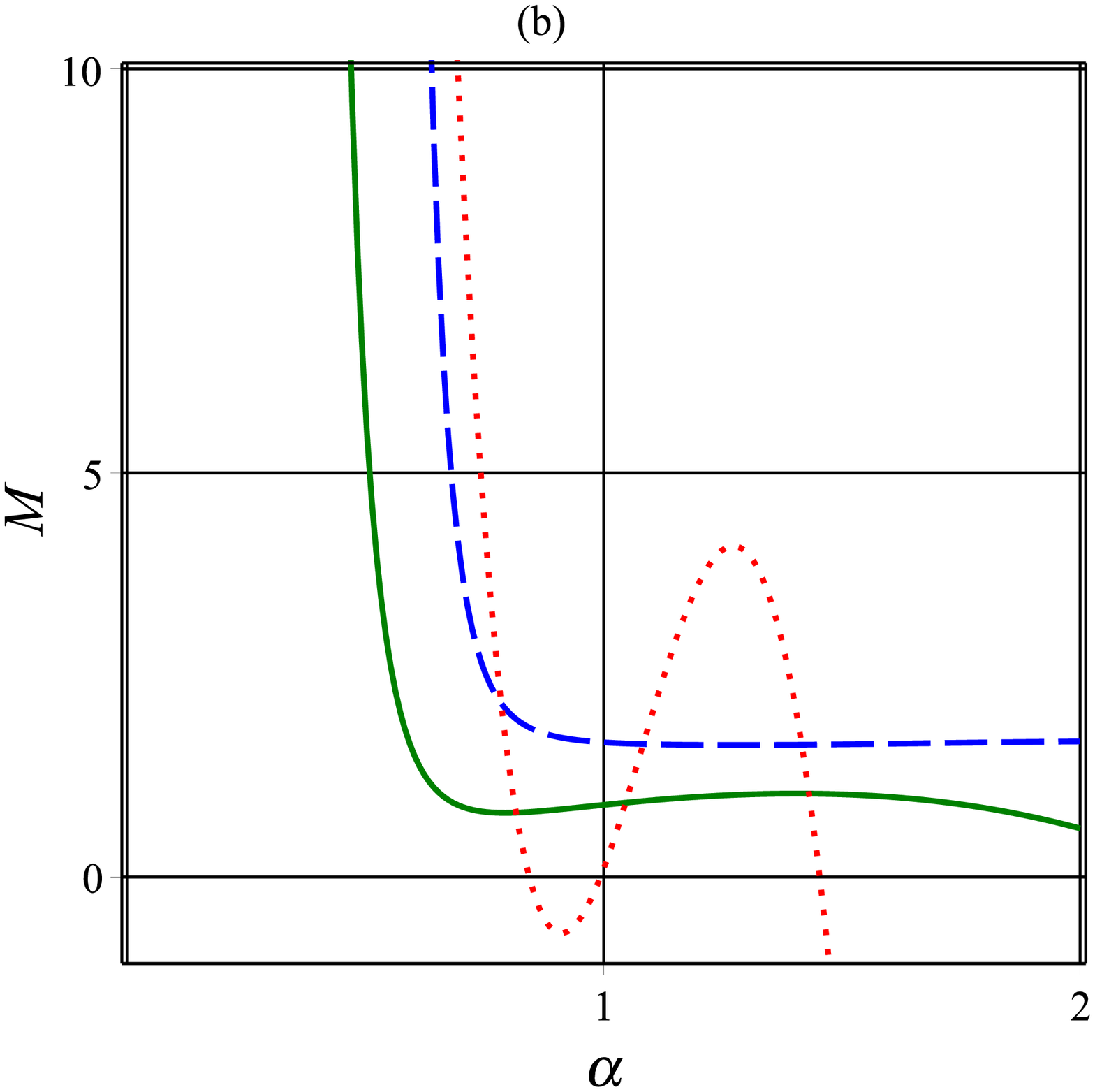}
 \end{array}$
 \end{center}
\caption{Typical behavior of $M$ in terms of (a) $r_{h}$ with $\alpha=1$, and (b) $\alpha$ with $r_{h}=1$,  by variation of the space-time dimensions. We set the black hole parameters as $a=b=l=\mu=1$, $C_l=C_M=2$. $d=4$ (blue dash), $d=5$ (green solid), $d=10$ (red dot).}
 \label{fig4}
\end{figure}

In this section, we come back to the parameters of the section 3 with specific $h$ function and investigate the effect of the logarithmic correction.
In order to study the effects of thermal fluctuations on thermodynamics, we consider ordinary space-time dimension  $d=4$, space-time with one extra dimension $d=5$ which is  important for AdS/CFT point of view  and space-time dimensions $d=10$ which corresponds to  superstring theory dimensions. Because of having some large expressions, we do graphical analysis of the model parameters.\\
In the Fig. \ref{fig2}, we draw temperature to see the effect of the thermal fluctuations and higher dimensions.
Here, it is clear that positive temperature is obtained only for special positive values of $\alpha$ (around $\alpha=1$). Also, the temperature is always negative for infinitesimal values of $r_h$ for space-time dimensions $d=4$ and $d=5$. However,  the positive temperature with respect to $r_h$ exists  only in case of space-time dimensions $d=10$. It means that at low dimensions the small black hole may be unstable and presence of $\alpha$ may help to gain stable black hole. In another word, right plot of Fig. \ref{fig2} shows that the AdS black hole with holographic dual of Van der Waals fluid is unstable without consideration of thermal fluctuations. Such instability happens when size of black hole is small (see left plot of Fig. \ref{fig2}) where thermal fluctuations is important. We will study stability of the black hole in the next section.\\
In the  plots of the Fig. \ref{fig3}, we can see behavior of the entropy under the effect of the thermal fluctuations and higher dimensions.
From the graph, it is clear that as long as the space-time dimension increases
the  entropy with respect to $r_h$ shows similar behavior with decreasing value towards negative. However, the   entropy with respect to $\alpha$ changes behavior for higher space-time dimensions $d=10$ which can be seen by red dot line.
This means that $\alpha$ affects  the entropy at higher dimensions much differently. It means that the AdS black hole with Van der Waals dual is almost unstable at 10 dimensions. On the other hand, at $d=4$ and $d=5$, the left plot of the Fig. \ref{fig3} shows that entropy is negative for $\alpha=1$ at $r_{h}<1$. It may be due to some instabilities corresponding to small black hole. It will be clear when we discuss stability of black hole by analyzing specific heat in the next section.\\
In the plots of the Fig. \ref{fig4} we can see behavior of the black hole mass under the effect of the thermal fluctuations and higher dimensions. We can see some negative regions in 10 dimensions which could be removed by suitable choice of $\alpha$. The  black hole mass
takes high value very sharply as $r_h$ tends to smaller values. For $d=10$, the entropy
from minimum value starts increasing with increasing $r_h$ but after a certain value
it starts falling again. In $d=4$ and $d=5$ we can see from the left panel of the Fig. \ref{fig4} that the black hole mass is increasing function of $r_{h}$ in allowed region ($r_{h}>1$) where the black hole is stable.\\

\begin{figure}[h!]
 \begin{center}$
 \begin{array}{cccc}
\includegraphics[width=60 mm]{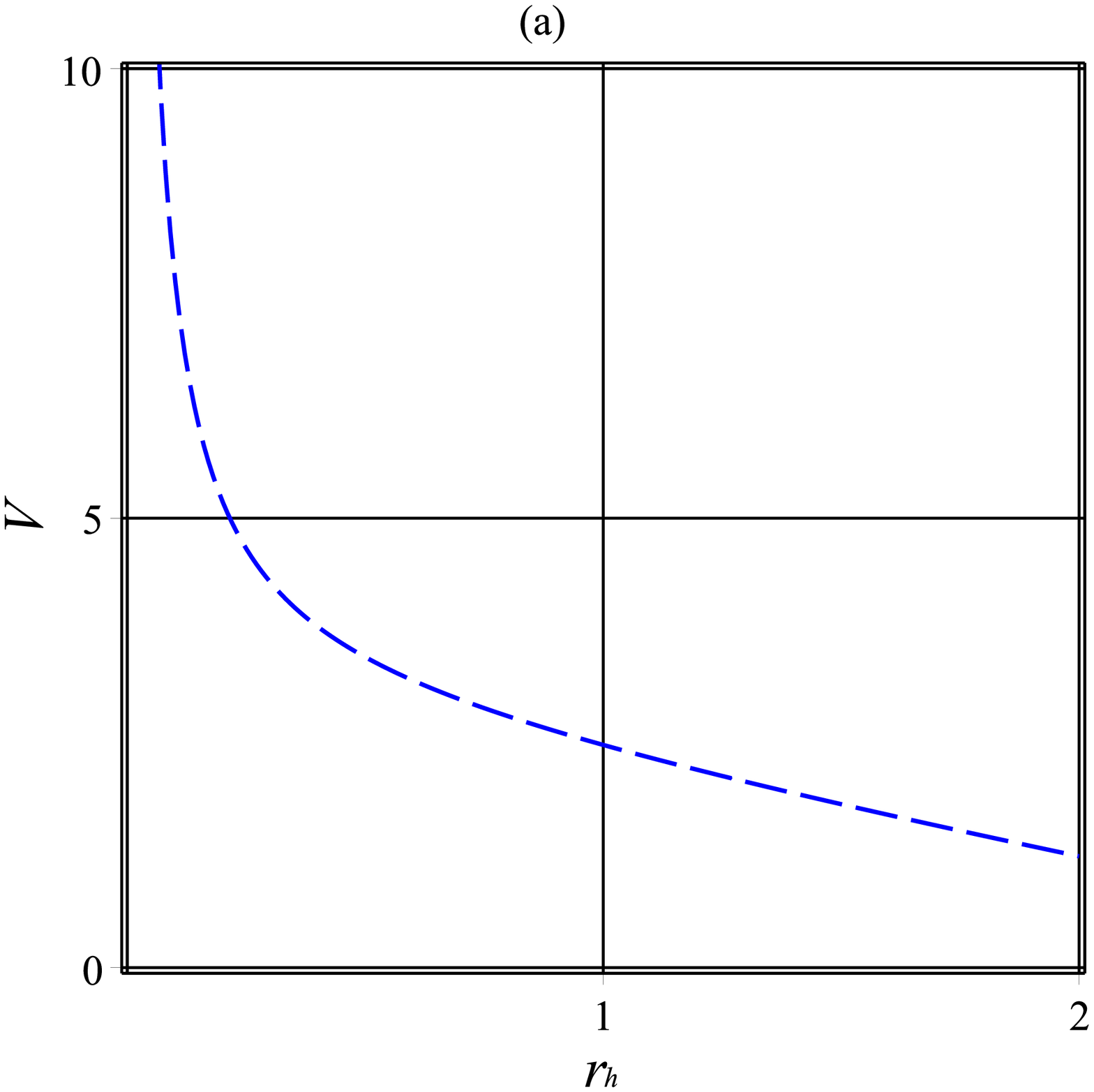}&\includegraphics[width=60 mm]{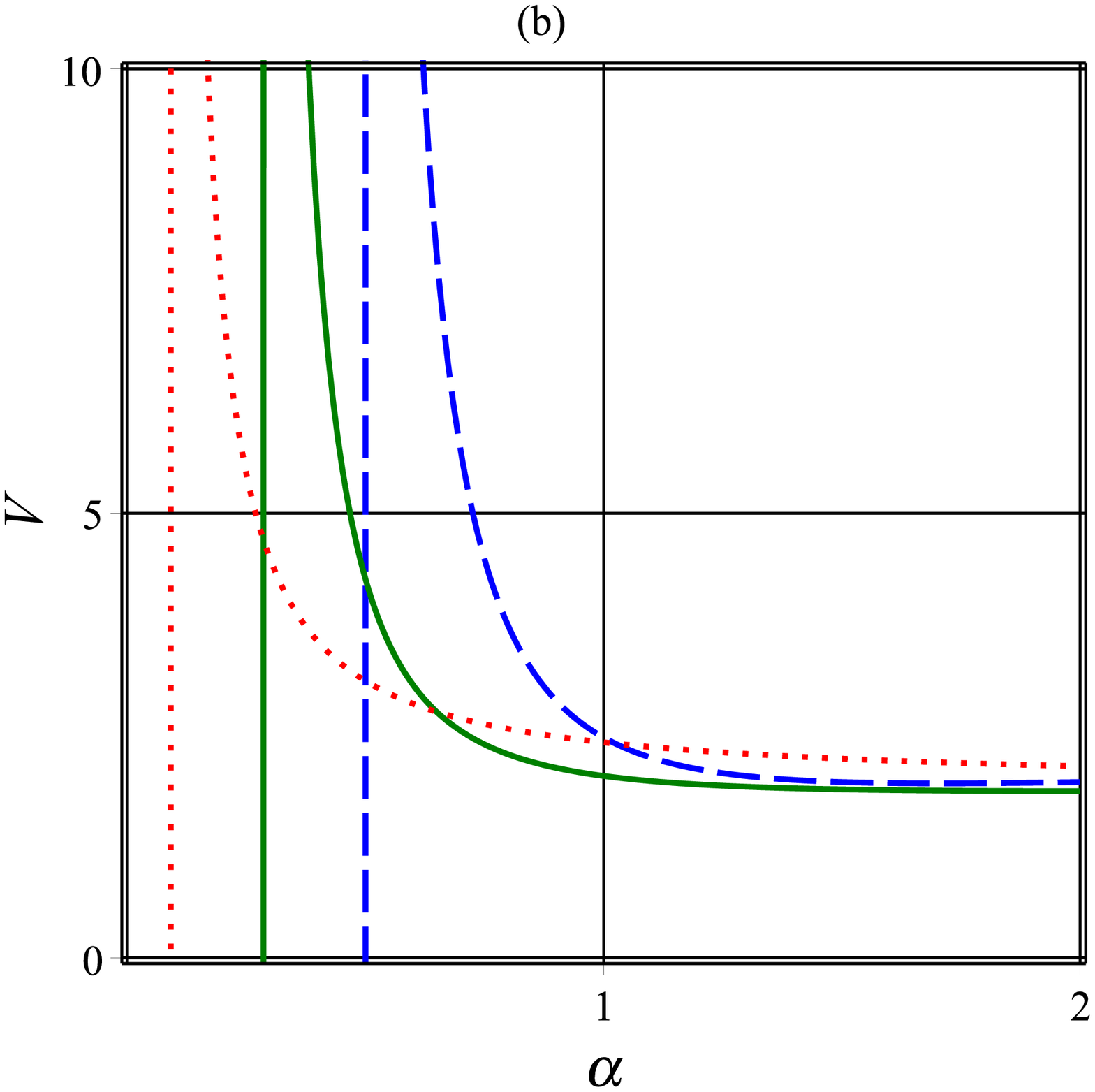}
 \end{array}$
 \end{center}
\caption{Typical behavior of the black hole volume in terms of (a) $r_{h}$ with $\alpha=1$, and (b) $\alpha$ with $r_{h}=1$,  by variation of the space-time dimensions. We set the black hole parameters as $b=1$, $C_l=2$. $d=4$ (blue dash), $d=5$ (green solid), $d=10$ (red dot).}
 \label{fig5}
\end{figure}

In the plots of the Fig. \ref{fig5}, we can see behavior of the black hole volume under the effect of the thermal fluctuations and higher dimensions. We can see the black hole volume is decreasing function of $\alpha$. It means that thermal fluctuations are important for the small black holes. By increasing size (volume) of black hole, logarithmic correction is small and negligible. Also, for $d=4$,
the black hole volume decreases as $r_h$ increases towards positive and becomes asymptotic
as $r_h$ tends to zero.\\

\begin{figure}[h!]
 \begin{center}$
 \begin{array}{cccc}
\includegraphics[width=60 mm]{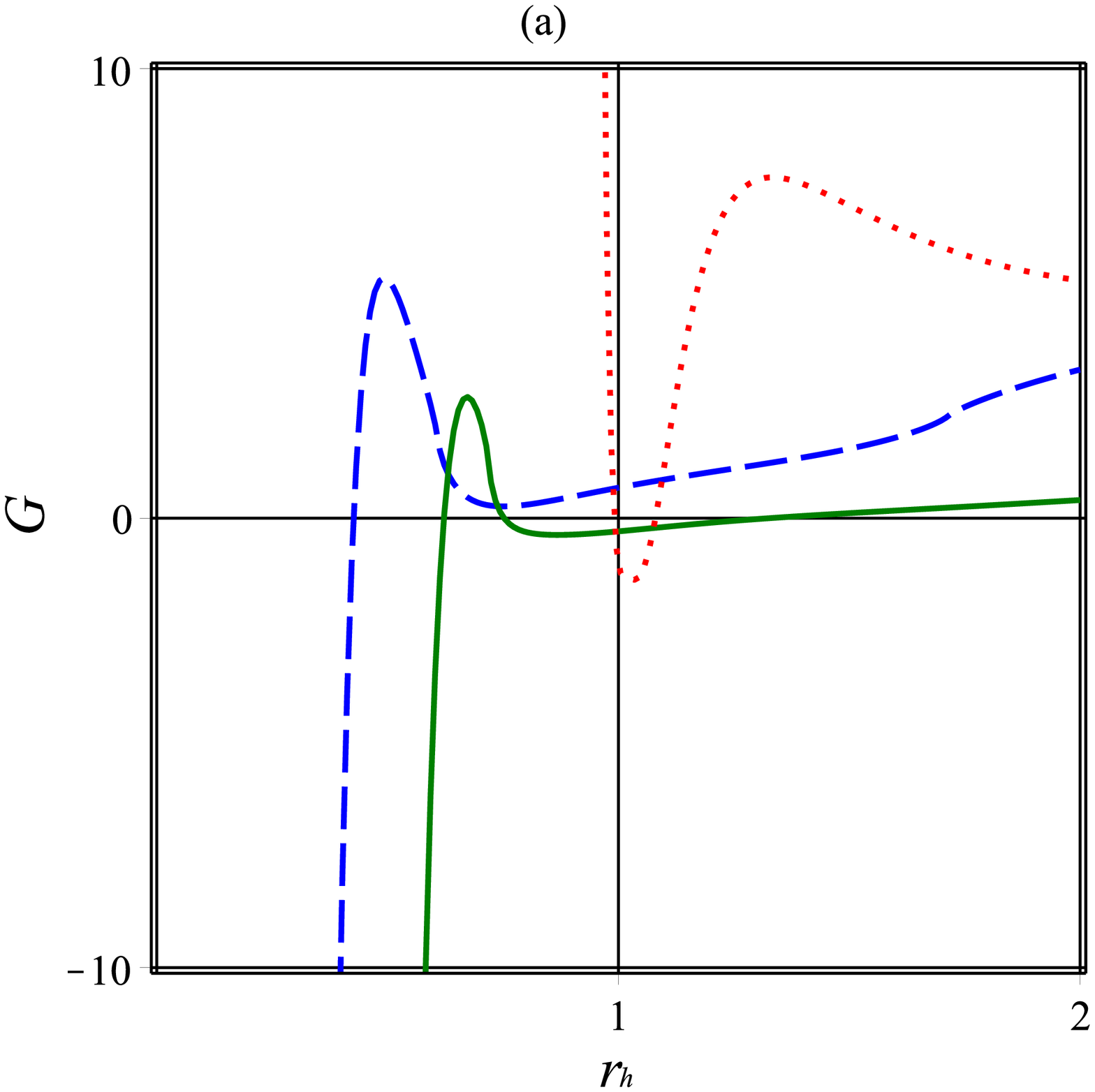}&\includegraphics[width=60 mm]{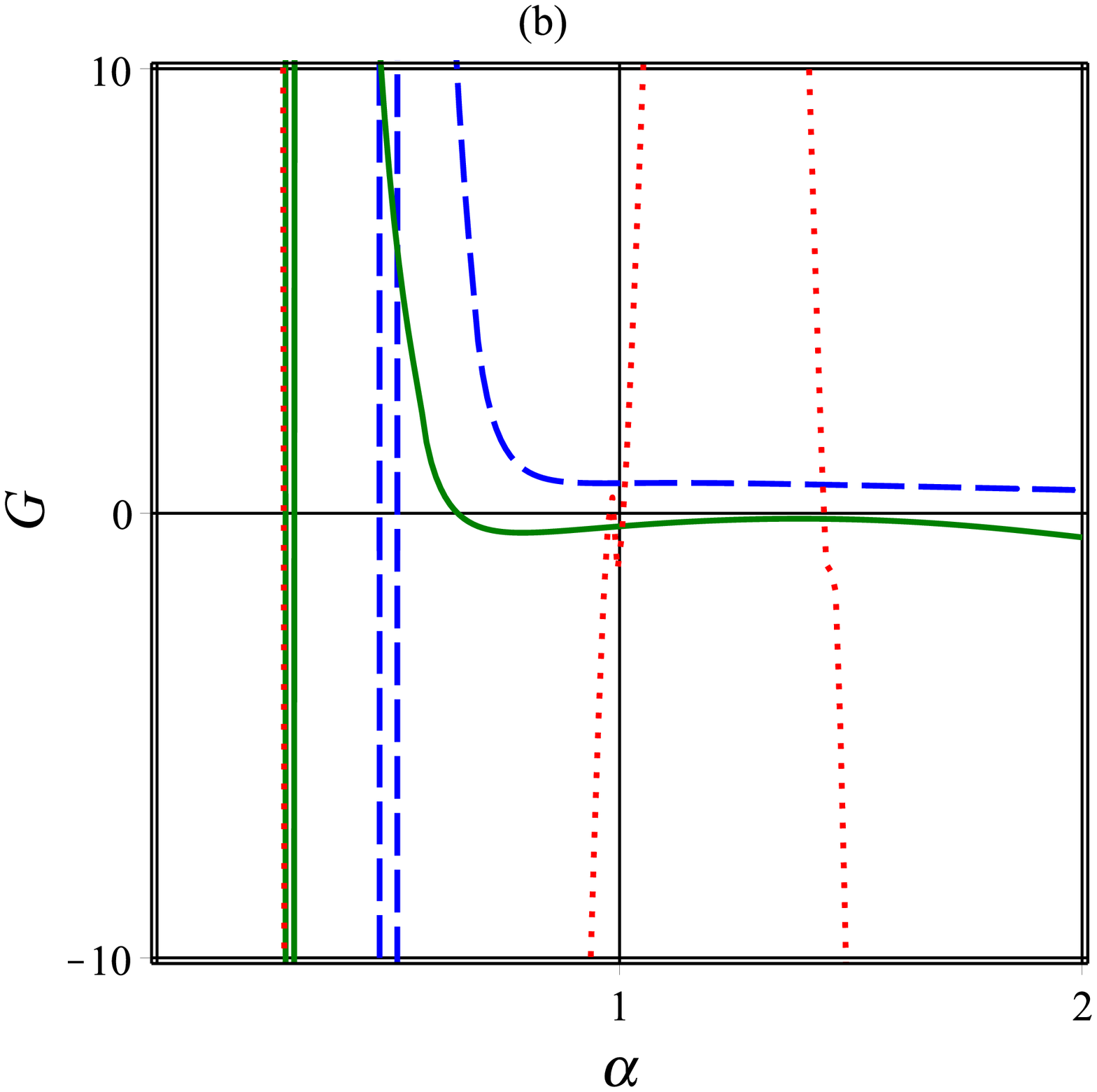}
 \end{array}$
 \end{center}
\caption{Typical behavior of Gibbs free energy in terms of (a) $r_{h}$ with $\alpha=1$, and (b) $\alpha$ with $r_{h}=1$,  by variation of the space-time dimensions. We set the black hole parameters as $a=b=l=\mu=1$, $C_l=C_M=2$. $d=4$ (blue dash), $d=5$ (green solid), $d=10$ (red dot).}
 \label{fig6}
\end{figure}

In the plots of the Fig. \ref{fig6} we can see behavior of the Gibbs free energy under the effect of the thermal fluctuations and higher dimensions. From the Fig. \ref{fig6}  (b), we can see a minimum for Gibbs free energy  concerning the thermal fluctuations for space-time dimensions $d=10$. However, the behavior is similar for the  space-time dimensions $d=4,5$ though the thermal fluctuations is effective for $d=5$ than $d=4$ for rather smaller value of $\alpha$. From the Fig. \ref{fig6}  (a), we see that the minimum value of
the Gibbs free energy with respect to $r_h$ is highly negative for space-time dimensions $d=4,5$, however for space-time dimensions $d=10$ this is not the case and the Gibbs free energy has (almost) positive values for all $r_h$.

\begin{figure}[h!]
 \begin{center}$
 \begin{array}{cccc}
\includegraphics[width=60 mm]{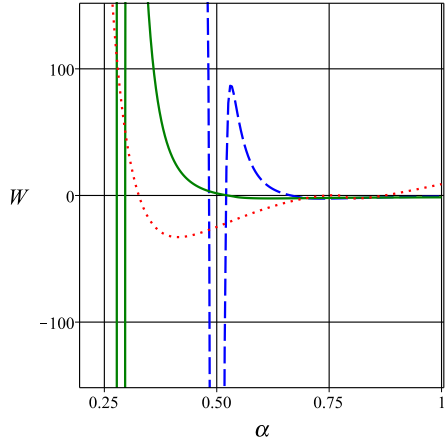}
 \end{array}$
 \end{center}
\caption{Typical behavior of $W=M-TS-PV$ in terms of $\alpha$ with $r_{h}=1$,  by variation of the space-time dimensions. We set the black hole parameters as $a=b=l=\mu=1$, $C_l=C_M=0$. $d=4$ (blue dash), $d=5$ (green solid), $d=10$ (red dot).}
 \label{figSmarr}
\end{figure}

Finally, we can investigate validity of the following relation,
\begin{equation}\label{Smarr}
M=TS+PV.
\end{equation}
In the Fig. \ref{figSmarr} we draw $W=M-TS-PV$ and shows that there are some zeros in the presence of the logarithmic correction. For example, we can see that in the case of $\alpha=1$, the equation (\ref{Smarr}) holds when $r_{h}\approx0.1$. As before, the situation for $d=10$ is a bit different and from dotted line we can see that the equation (\ref{Smarr}) formula holds for $\alpha\approx0.8$.

\section{Stability}
\begin{figure}[h!]
 \begin{center}$
 \begin{array}{cccc}
\includegraphics[width=50 mm]{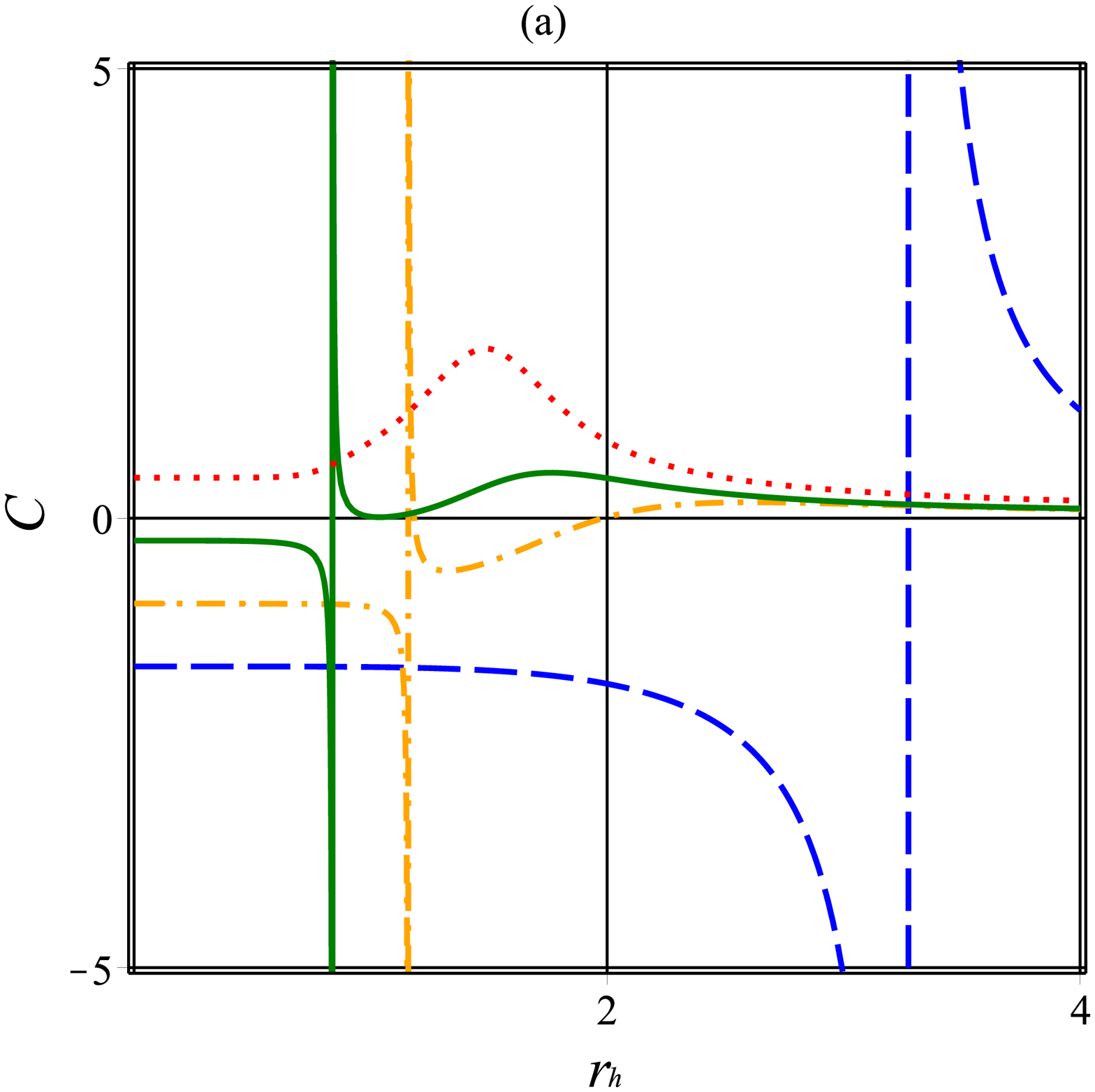}&\includegraphics[width=50 mm]{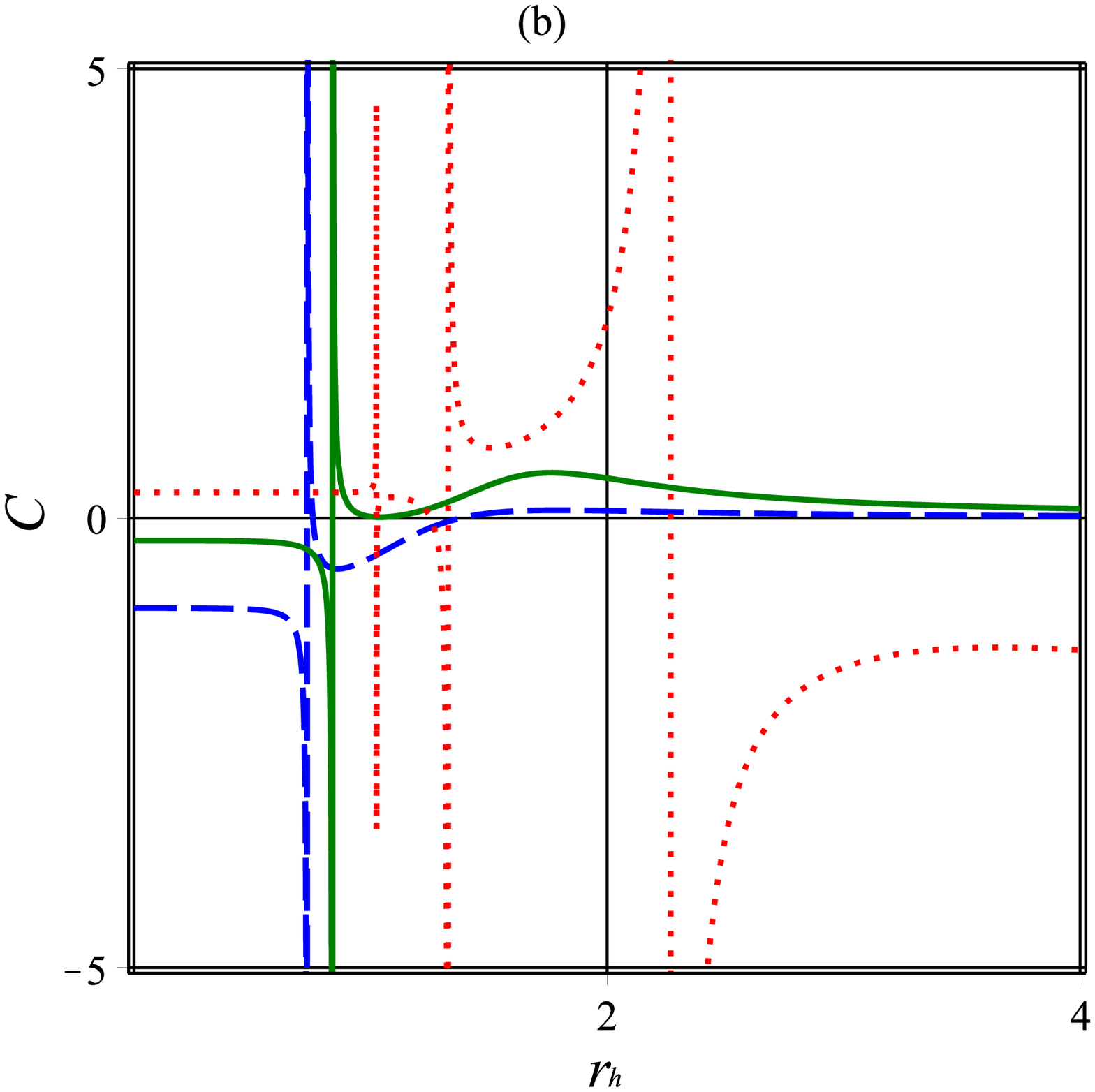}&\includegraphics[width=50 mm]{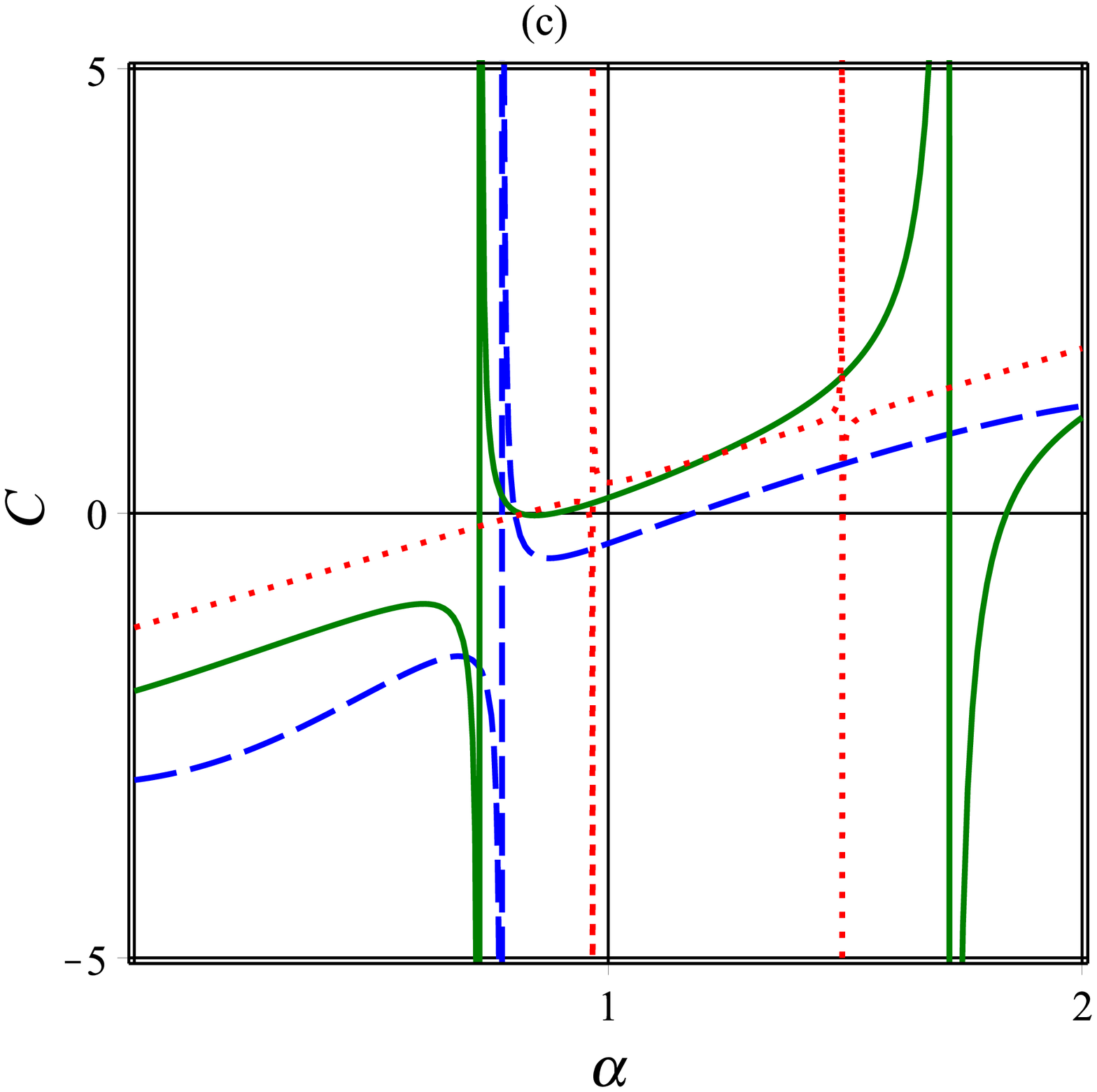}
 \end{array}$
 \end{center}
\caption{Typical behavior of the specific heat in terms of $r_{h}$ and $\alpha$ by variation of the black hole parameters with $a=b=l=1$ and $\mu=C_l=C_M=2$. (a) $d=5$; $\alpha=0.2$ (blue dash), $\alpha=0.6$ (orange dash dot), $\alpha=1$ (green solid), $\alpha=1.4$ (red dot). (b) $\alpha=1$; $d=4$ (blue dash), $d=5$ (green solid), $d=10$ (red dot). (c) $r_{h}=1$; $d=4$ (blue dash), $d=5$ (green solid), $d=10$ (red dot).}
 \label{fig7}
\end{figure}

Now, we can study stability of the model using the equation (\ref{C}). It seems enough to establish stability of black hole because in absence of chemical potential, Hessian of Helmholtz free energy is not important and we need only to find sign of the specific heat. In the Fig. \ref{fig7}, we draw specific heat in terms of $r_{h}$ in (a) and (b) plots and in terms of $\alpha$ in (c) plot. In the Fig. \ref{fig7} (a) we consider the case of $d=5$ and see that small $\alpha$ includes some unstable regions. Almost of these regions   correspond  to negative entropy which is discussed in previous section. Then, by increasing $\alpha$ domain of instability reduced, so for the case of $\alpha=1.4$ we have completely stable black hole. In this case we can see a peak for the specific heat which is like Schottky anomaly, and hence we can have an excitation gap above the ground state.

In figure  \ref{fig7} (b), we see the behavior of specific heat in terms of $r_h$
with fixed thermal fluctuation and varying space-time dimensions. Here,  we observe that the instability  increases
 as space-time dimensions increase with various minima, which can be clearly seen from the plot for $d=10$. The plot  \ref{fig7} (c) shows the behavior of specific heat with respect to
 $\alpha$ with fixed $r_h$ and varying space-time dimensions. Remarkably, we found that   for $d=5$
 the the instability of specific heat occurs again when $\alpha$ increases slightly or decreases slightly value. However, for $d=4$ the stability exists for larger span of $\alpha$.
 For $d=10$, the more instability occurs of different character with respect to lower
 dimensions.
\section{Conclusions}
In this paper, we have analyzed the effects of the logarithmic correction to the thermodynamics of the higher dimensional AdS black hole. This logarithmic correction comes from the leading-order thermal fluctuations of black hole. In this context, we have given the basic set-up for the higher dimensional AdS black hole  described by a suitable metric in terms of an unknown function $h(r, P_\Lambda)$
which can be determined with the help of the thermodynamics relations. Furthermore, we
have computed various thermodynamics quantities describing more exact equations of state
due to the first-order (logarithmic) correction in entropy of the system. For instance,
we have derived the first-order corrected  enthalpy, volume, Gibbs free
energy, Helmholtz free energy and specific heat.
Eventually, we have verified the first law of thermodynamics for the first-order corrected
higher dimensional black hole.

Furthermore, in order to have specific form of the unknown function $h(r, P_\Lambda)$,
we have considered the logarithmic corrected  black hole equation of state
and compared it with the corresponding
fluid with Van der Waals equation of state, while we identify the black hole
and fluid temperatures, the black hole and fluid
volumes, and the cosmological and fluid pressures. Here, we have found a
typical behavior of the function $h$ in terms of the model parameters. For instance,
  the function $h$ increases sharply with respect to the small
$r_h$ and becomes saturated for larger (of the order of unit) $r_h$ as can be seen in Fig.  \ref{fig1}. We have analyzed the effects of thermal fluctuations on various
equation of states. Namely, the effects of thermal fluctuations are discussed for
temperature, entropy, mass, volume and Gibbs free energy.  Finally, we have discussed the
stability of the model for different higher space-time dimensions, namely, $d=4, 5$ and $d=10$.
These particular dimensions are considered due to  their physical importance. These particular dimensions
are considered due to their physical importance. In the dual QFT, the van der Waals
(VdW) black hole corresponds to a thermal state with VdW EoS. The five dimensional
black hole may be important from AdS/CFT point of view, where a five dimensional
black hole background has holographic dual CFT. Finally, 10-dimensional black hole may
be important from superstring theory. In all above, one can use logarithmic corrected
thermodynamics to study quantum gravity effects.  In summary, we have shown that the presence of logarithmic correction affects strongly stability of black hole. It is easy to extend our work to the case of charged AdS black hole in higher dimensional space-time. Also one can consider higher order corrections of the entropy to investigate thermodynamics quantities. These are subjects of our future works.

\end{document}